\documentclass[structabstract]{aa}  
\usepackage{graphicx}
\usepackage{txfonts}
\usepackage{natbib}
\usepackage{indentfirst}
\usepackage{url}
\usepackage{color}

\newcommand{\hho}  {H$_2$O }
\newcommand{\mJybm}{mJy~beam$^{-1}$}
\newcommand{\Jybm}{Jy~beam$^{-1}$}
\newcommand{\vlsr}{V$_{\rm LSR}$ }
\newcommand{\kms} {km~s$^{-1}$}
\newcommand{\cmt} {cm$^{-3}$}
\begin{document}
   \title{The magnetic field of IRAS 16293-2422 as traced by shock-induced \hho masers}

   \author{F. O. Alves\inst{1}, W. H. T. Vlemmings\inst{2}, J. M. Girart\inst{3} \and J. M. Torrelles\inst{4}}
          
   \institute{Argelander-Institut f\"ur Astronomie, University of Bonn, 
             Auf dem H\"ugel 71, D-53121 Bonn, Germany \\
             \email{falves@astro.uni-bonn.de}
         \and
              Chalmers University of Technology, Onsala Space Observatory, SE-439 92 
              Onsala, Sweden \\
              \email{wouter.vlemmings@chalmers.se}
         \and
              Institut de Ci\`encies de l'Espai (IEEC--CSIC), Campus UAB,
              Facultat de Ci\`encies, C5 par 2$^{{\mathrm a}}$, 08193 Bellaterra, Catalunya, Spain\\
              \email{girart@ice.cat}
	\and
	    Institut de Ci\`encies de l'Espai (CSIC)-UB/IEEC, Universitat 
	    de Barcelona, Mart\'i i Franqu\`es 1, E-08028 Barcelona, Spain \\
	    \email{torrelles@ieec.cat}
                        }

   \date{Received ; accepted }

 
  \abstract
   {Shock-induced \hho masers are important magnetic field tracers at very high density gas. 
   Water masers are found in both high- and low-mass star-forming regions, acting as a powerful tool
    to compare magnetic field morphologies in both mass regimes.}  
   {In this paper, we show one of the first magnetic field determinations in the low-mass protostellar core IRAS 
   16293-2422 at volume densities as high as 10$^{8-10}$ \cmt. Our goal is to discern if the collapsing regime
   of this source is controlled by magnetic fields or other factors like turbulence.}
   {We used the Very Large Array (VLA) to carry out spectro-polarimetric observations in the 
   22 GHz Zeeman emission of \hho masers. From the Stokes V line profile, we can estimate the magnetic field 
   strength in the dense regions around the protostar.}
   {A blend of at least three maser features can be inferred from our relatively high spatial resolution data set
   ($\sim$ 0.1$^{\prime\prime}$), which is 
   reproduced in a clear non-Gaussian line profile. The emission is very stable in polarization fraction 
   and position angle across the channels. The maser spots are aligned with some components of the complex
   outflow configuration of IRAS 16293-2422, and they are excited in zones of compressed gas produced
   by shocks. The post-shock particle density is in the range of $1-3 \times 10^9$ \cmt, consistent with typical
   water masers pumping densities. Zeeman emission is produced by a 
   very strong line-of-sight magnetic field ($B \sim 113$ mG) }
   {The magnetic field pressure derived from our data is comparable to the ram pressure of the outflow
   dynamics. This indicates that the magnetic field is energetically important in the dynamical evolution of IRAS 16293-2422.}
   

   \maketitle
%

\section{Introduction}
\label{intro_sec}

Spectro-polarimetric observations of masers are a powerful tool to study the magnetic field properties  in 
the maser pumping zone. Water masers are unique because they are found in a variety of astrophysical 
environments and, in particular, in star-forming regions at distinct mass regimes. 
In contrast, methanol and OH masers are found associated mostly with high-mass star formation sites 
\citep[e.g.,][]{Vlemmings08,Sanna10a,Sanna10b}. 
The most commonly observed water maser line is the  (6$_{16} - 5_{23}$) transition at 22 GHz, an excellent probe 
of molecular gas at very high volume densities 
\cite[$n_{\rm H_2}$ in the $10^{8}$ to $10^{10}$ cm$^{-3}$ range,][]{Elitzur89}. 
While the Zeeman splitting is small ($\sim 10^{-3}$ Hz $\mu$G$^{-1}$),  
the line strength is sometimes sufficient for measuring the circular polarization and therefore for a direct 
measurement of the line-of-sight (LOS) component of the magnetic field. If linear 
polarization is also measured, then the full 3-D magnetic field configuration can be derived. 

Recently, several studies have appeared in the literature reporting the efficiency of maser spectro-polarimetry 
technique using masers. These studies have revealed the magnetic field properties in very dense molecular 
environment around the circumstellar envelopes of evolved stars and young stars 
\citep[e.g.,][]{Surcis11a,Surcis11b,Perez11}.  
A representative result regarding how magnetic fields can be resolved in very small spatial scales was obtained 
by \citet{Vlemmings06b}, who carried out Very Long Baseline Array (VLBA) polarization observations of  \hho 
masers around the Cepheus A HW2 high-mass object. 

Thermal dust polarization emission is difficult to observe in the most embedded 
portions of molecular clouds due to the low sensitivity to polarized data (and the feasibility is then limited toward a 
handful list of objects). 
Submillimeter (submm) emission suffers from 
depolarization effects like unresolved fields or roundness of dust grains 
\citep{Goodman92,Goodman95,Lazarian97}. Water masers can be thought as an observational tool to overcome
this issue, since they are excited 
at very high densities. 

IRAS 16293-2422 (hereafter, I16293), is a 
prototypical low-mass protostellar system located in the $\rho$ Ophiuchus molecular cloud 
\citep[$d \simeq 120$ pc,][]{Loinard08}. 
This source is a well-studied binary system, usually referred as source A and B, with angular separation of 
$5^{\prime\prime}$ (600 AU) and very embedded in a dense molecular core \citep{Wootten89,Looney00}. Both 
sources have a very rich chemistry which is typically found in hot cores 
\citep{Ceccarelli00,Kuan04,Bisschop08,Jorgensen11}. 
Although component B appears to be a single source,
high resolution interferometer data reveal a higher degree of fragmentation for source A.
In fact, two dust submm components were detected within this object
with an angular separation of $\sim 0.6^{\prime\prime}$ 
\citep[$\sim 72$ AU,][]{Chandler05}. I16293 has two large scale ($\sim$ 1$^\prime$) bipolar CO outflows, one 
of them associated with source A while the powering source of the other  CO outflow is a matter of debate 
\citep{Walker88,Stark04,Yeh08}. Recently, observations of the SiO (8--7) emission have revealed a compact 
molecular outflow also associated with source A \citep{Rao09}. I16293 has strong water maser emission that has 
been well monitored {\citep{Wilking87,Terebey92,Claussen96,Furuya03}.
The strongest features appear at blueshifted and redshifted LSR velocities with respect to the ambient
cloud velocity ($\sim$ 4 \kms), typically between  
\vlsr of -5 and 10 \kms. The \hho maser emission very often has intensities of more than 100 Jy, and at 
times it is stronger than 300 Jy, at LSR velocities of $\sim 7$ \kms.

\citet{Tamura93} performed observations of the 1.1 mm dust polarized emission toward I16293 at an angular 
resolution of 19\arcsec. They found that the magnetic field lines  are perpendicular to the major axis of the dense 
elongated disk-like molecular structure ($\sim 10^{\prime\prime}$ size, $\sim$ 1200 AU) observed in C$^{18}$O by 
\citet{Mundy90}. This configuration was corroborated recently by \citet{Rao09},  who used the Submillimeter
Array (SMA) and obtained 
a polarization map at an angular resolution of $\sim 2''$ ($\sim 240$~AU), resolving both sources A and B. The 
mean volume density traced by the SMA maps is $\sim 6 \times 10^{7}$~\cmt. The polarization pattern around 
source A  is compatible with a hourglass morphology for the magnetic field, whose strength was estimated in 
$\sim$ 4.5~mG. The SMA maps show that there is a misalignment between the outflow direction and 
the magnetic field axis, roughly in agreement with model predictions where  the magnetic energy is 
comparable to the centrifugal energy.  In contrast, source B is associated with an uniform and apparently 
undisturbed magnetic field. 
 
In this work, we report spectro-polarimetric Very Large Array (VLA) observations of \hho masers toward 
I16293. In \S~\ref{sec_obs_mas}, we describe the details of the observational setup. In \S~\ref{sec_res_mas}, 
the results obtained from the maser spectroscopy are shown. In \S~\ref{mas_outf_sec}, we discuss a possible 
correlation between the masers features and the outflows. The magnetic field strength estimation and 
its implication on the core dynamical evolution are discussed in sections \ref{field_stre_sec} and 
\ref{core_evolution}, respectively. Finally in \S~\ref{sec_mas_con} we summarize our conclusions.

\section{Observations}
\label{sec_obs_mas}

The observations were done with the VLA (NRAO\footnote[1]{The National Radio Astronomy 
Observatory - NRAO - is a facility of the National
Science Foundation, operated under cooperative agreement by Associated Universities, Inc.}, 
New Mexico, USA) in its extended 
A-configuration, on 2007 June 25$^{\mathrm{th}}$ and 27$^{\mathrm{th}}$. The tracks lasted $\sim$ 5.5 hours 
each. A total of 27 antennas were used, 10 of them already retrofitted with the new system, resulting in a 
combined VLA/EVLA (Extended VLA) observation. We used the K band receivers (22-24 GHz) tuned 
at the frequency of the water maser (6$_{16} - 5_{23}$)
rotational transition ($\nu_0 = 22.23508$ GHz). We used the full polarization capability of the 
correlator, selecting a bandwidth of 0.7813 MHz ($\sim 10.5$~\kms\ in velocity). The spectral setup contains a total 
of 128 channels which covers most of the velocity range of the strongest water maser features (observed 
around the brightest feature previously reported at $\sim 7$ \kms)
at the spectral resolution of 0.08~\kms. The quasar 
J1626-298 was used as gain calibrator. The quasar J1331+305 and the radio source J1751+096 were used for 
polarization calibrations in order to obtain corrections in the instrumental feed polarization and the corrected 
position angle of the polarization vectors. All three calibrators were also used for bandpass corrections. We 
performed self--calibration using the channel with the strongest intensity. Since the detected emission is 
unresolved, both phase and amplitude self--calibration solutions were applied to the other channels.

The combined VLA/EVLA setup introduces about 8\% closure errors on VLA-EVLA baselines-based 
corrections on the bandpass. This happens due to the distinct bandpass response of the EVLA antennas compared 
to the VLA ones. These errors become more severe at narrow bandwidths, when the processing of the
digital signal of EVLA antennas aliases the power response at the band edges. 
During the data reduction, we took special care on determining the bandpass solutions by performing particular
calibration strategies recommended to the user in the NRAO 
webpage\footnote[2]{\url{http://www.vla.nrao.edu/astro/guides/evlareturn/postproc/index.shtml#closure-line}}. 
They consisted of flagging EVLA-EVLA baselines prior to bandpass calibration and unflagging them prior to 
applying the bandpass solutions to the data set. The bandpass corrections are then
applied to an averaged single-channel multi-source dataset (``channel 0'' file) and to the 
multi-channel original file.

Data reduction was done with the Astronomical Image Processing Software package (AIPS). Imaging of Stokes 
parameters I, Q, and U were generated with a quasi-uniform weighting \citep[{\it robust} = -1, ][]{Briggs95}. Maps of polarized fraction 
($P$) and position of polarization angles (PA) were obtained by combining Stokes Q and U images in such a way 
that $P = \frac{I_{P}}{I} = \frac{\sqrt{Q^2 + U^2}}{I}$ and PA $= \frac{1}{2} \tan^{-1}(\frac{U}{Q})$. The resulting 
synthesized beam is $0.14'' \times 0.08''$, with a position angle of $-5.7\degr$. The rms noise for channels 
where no emission is detected is $\sim  8$ \mJybm, and it increases up to 23 \mJybm\, at the peak emission 
channel. A slightly lower rms is  observed for polarized intensity. 

\section{Results}
\label{sec_res_mas}

\subsection{\hho maser line}
\label{water_line_subsec}

The contour channel maps of the \hho emission observed with the VLA/EVLA data are shown in Fig.  
\ref{water_chan}. 
The emission extends over a wide range of velocities ($4.5 <$ \vlsr$ < 9$ 
\kms), i. e., at redshifted LSR velocities with respect to the cloud systemic velocity ($\sim 4$ \kms). 
Our spectral setup covers only a small portion of LSR velocities blueward of it ($\sim 2 < $ \vlsr $< 4$ 
\kms) but no emission was detected within this spectral range.
The water maser line has peak intensity of $170$ \Jybm, detected at \vlsr $\simeq 7.4$ \kms. 
This emission
was extensively reported in previous surveys also as the brightest feature \citep[see, for instance,][]{Claussen96,Furuya03}.
Some channels 
have emission extended slightly eastward (\vlsr between $\simeq 6.5$ and $7$~\kms), suggesting
that the observed emission in these channels is marginally resolved.

\begin{figure*}
\centering
\includegraphics[angle=-90,width=16cm]{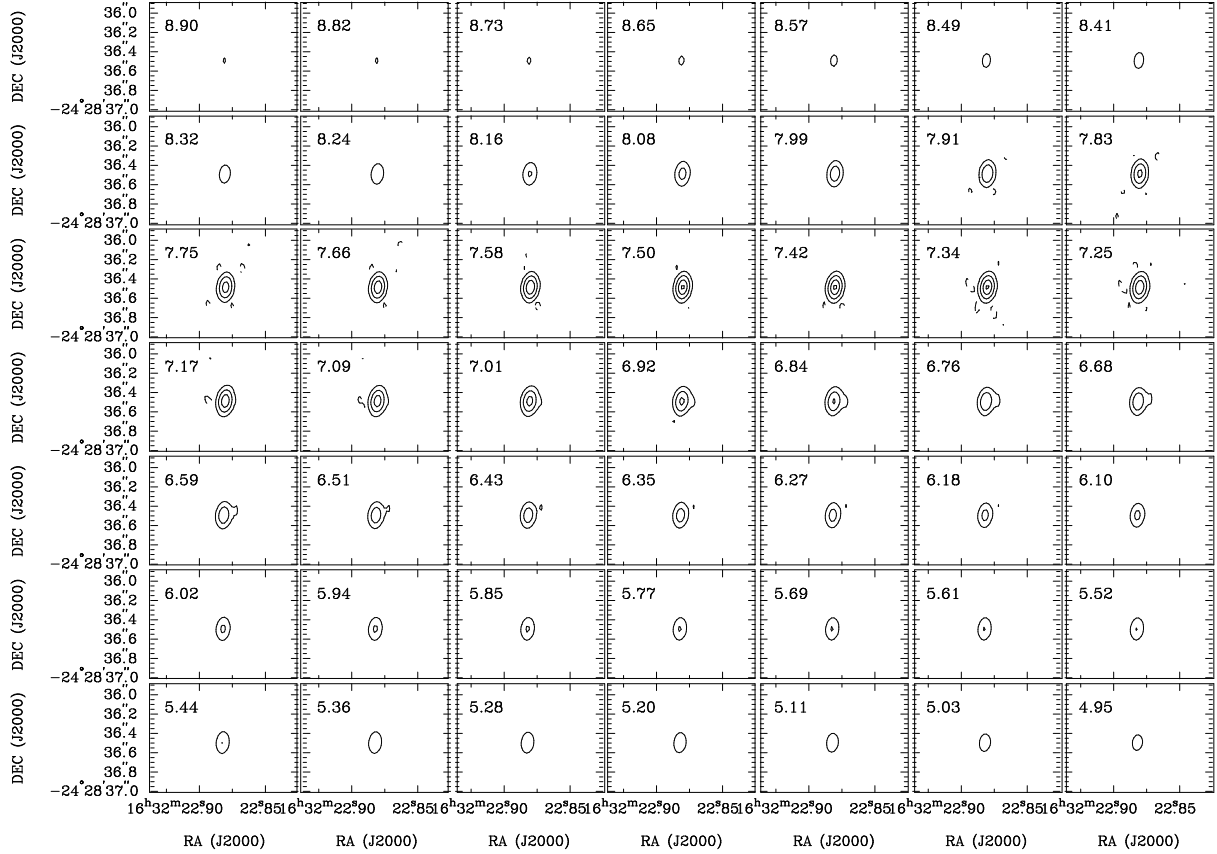}
\caption{Channel maps of the \hho emission toward IRAS 16293-2422. 
Contours are -50, 50, 500, 3$\times$10$^{3}$, 1$\times$10$^{4}$, 2$\times$10$^{4}$ times 8~mJy~beam$^{-1}$, 
the rms noise of the maps with lower intensities (see Section \ref{sec_obs_mas}).
The peak flux of 167 \Jybm\ occurs at \vlsr = 7.4~\kms. The LSR velocity of each channel 
is labelled.} 
\label{water_chan}
\end{figure*}

A scheme of the distribution of dust and molecular outflows in I16293 is shown in Fig.  \ref{dust_dist}. The 
masers detected in our observations are associated with source A. They are located $\sim 0.25$\arcsec\, ($\sim 
30$ AU in projection) to the southeast of the dust condensation $Aa$ resolved by the subarcsecond 
submillimeter (submm) observations of \citet{Chandler05}. 

The Stokes I spectrum of the maser emission is shown in Fig.  \ref{stokesIpolfrac} (upper panel).  
The non-Gaussian 
line profile indicates that there are unresolved components in the spectrum. Apart 
of the peak intensity at 7.4 \kms, unresolved emission seems to be also 
present at lower velocities (\vlsr $\simeq$ 5.5 
\kms) with a strong flux of $\sim$ 20 \Jybm. Fainter emission ($\sim 5$~\Jybm) 
is also observed at higher velocity channels (\vlsr $\simeq$ 9.2~\kms). Therefore, there are at least three 
unresolved components (Table \ref{maser_feat_tab}). 
A two-dimensional Gaussian profile fit on each of those components provides a mean spatial separation of 
$\sim 22$~milli-arcseconds (mas), which is higher than the precision on the relative position determination 
(rms$_{pos} \simeq 2$ mas, 
estimated from the ratio between the width of the synthesized beam and 
the signal-to-noise ratio of the fainter component).

\subsection{Polarized emission}
\label{sec_polem}

The spectrum of linearly polarized intensity is very similar to the Stokes I line profile 
except for the flux scale. It peaks at the same systemic velocity as the Stokes I spectrum, 
although it is weaker by a factor of $\sim 30$.
The dependence of the linear polarization intensity, polarization fraction, position angle and Stokes I with systematic 
velocity is shown in Fig. \ref{stokesIpolfrac}. The measured polarization fraction is $2.5 \pm 0.2\%$.  The 
polarization position angle $\theta$ is $-23 \degr$ and shows only small changes across the maser ($
\sigma_{\theta} = 2\degr$), implying that the polarization vectors at different velocities trace basically the same 
region.  The formal uncertainty 
in PA \citep[$\sigma_{\theta} = \frac{1}{2}\frac{\sigma_{P}}{I_{P}} \frac{180\degr}{\pi}$,][]{Wardle74} is small and
ranges between $\sim$ 0.14$\degr$ and 0.80$\degr$ considering channels with strong and weak linear 
polarization emission, respectively.
The linear polarization as derived from Stokes U and Q maps is exhibited in Fig.  \ref{maser_pol}, which shows 
the distribution of polarization vectors in the brightest velocity channel. The peak of polarized intensity is offset 
0.05\arcsec\,with respect to the Stokes I peak. The polarization vectors can be parallel or perpendicular to magnetic 
field orientation in the plane of the sky. In \S~\ref{field_stre_sec}, we discuss which assumptions must be 
considered in order to solve this ambiguity. 

\begin{table}[t]
\caption{\hho maser components in IRAS 16293-2422\label{maser_feat_tab}}
\centering
\tabcolsep 0.3cm
\begin{tabular}{c c c c}
\hline\hline
\vlsr & I$_{peak}^{\mathrm{a}}$ & $\alpha$ (2000) & $\delta$ (2000)  \\
(\kms) & (\Jybm) & (${\mathrm h}$ ${\mathrm m}$ ${\mathrm s}$) & (o $\prime$ $\prime\prime$) \\
 \hline
5.7 & 23 & 16 32 22.8830 & -24 28 36.495 \\
7.4 &  168 & 16 32 22.8808 & -24 28 36.487 \\
9.2 & 5 & 16 32 22.8818 & -24 28 36.493 \\
\hline
\end{tabular}
\begin{list}{}{}
\item[$^{\mathrm{a}}$]  Equatorial coordinates derived with the JMFIT task of AIPS.
\end{list}
\end{table}

The line profile of the circular polarization (Stokes V) has 
the characteristic S-shape (see Fig.~\ref{stokesV}, lower panel). The 
Stokes V spectrum is proportional to the first derivative of the Stokes I spectrum and to the LOS component 
of the magnetic field. The red dashed line in Fig. \ref{stokesV} indicates the fit which best represents this 
derivation. Due to remaining gain differences among the two polarizations, scaled down replicas of Stokes I
spectrum might contaminate the Stokes V data \citep{Sarma01,Vlemmings02}. These features were considered 
in the fit and subtracted from the synthetic V spectrum. 
The fraction of circular polarization, calculated as $(V_{max} - V_{min})/I_{max}$, is $\sim 
0.45$\% for the brightest component. The remaining maser features show only residual Zeeman profiles with 
amplitudes at the rms level. 

\section{\hho masers and the SiO/CO outflows}
\label{mas_outf_sec}

Fig.  \ref{maser_pos} shows that the three possible maser features unresolved in our data are distributed 
linearly. The velocity shift observed between \vlsr $\simeq 5.7$ \kms\,and higher velocities is 
oriented in a
E--W direction (PA $\simeq 110\degr$). The proximity with source $Aa$ and the complex outflow configuration
observed in this zone suggest that the maser emission is being pumped in the dense circumstellar material 
around this object. 

\begin{figure}[t]
\centering
\includegraphics[width=\columnwidth]{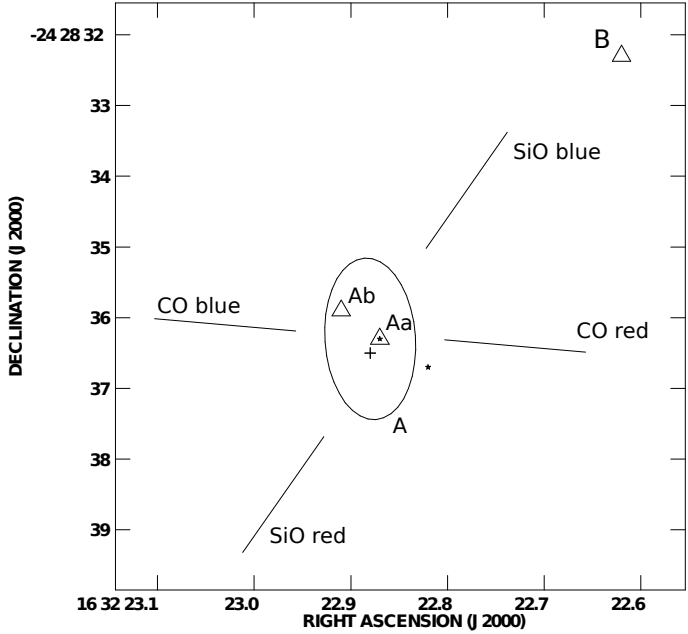}
\caption{Scheme of the distribution of dust and molecular material in I16293. The plus signal 
indicates the position of the peak intensity of our VLA water maser data. The ellipse is the deconvolved 
size of the dust continuum source A as derived by \citet{Rao09}. Triangles denote the position of the 
submillimeter condensations observed by \citet{Chandler05}. Stars denote the VLBI water 
maser detections of \citet{Imai07} and straight lines denotes the direction of the CO and 
SiO outflows associated with this core \citep{Rao09}.}
\label{dust_dist}
\end{figure}

\citet{Rao09} report a strong and very young SiO outflow extending toward the northwest-southeast (NW--SE) 
direction, with the SE lobe detected at redshifted LSR velocities (see Fig. \ref{dust_dist}). 
This SE component is also traced by a CO (3-2) outflow at \vlsr $\simeq 9$ \kms\, \citep[Fig. 7 in][]{Rao09}. The
CO outflow has a dominant E--W orientation and has already been reported in the literature \citep{Yeh08}.
\citet{Rao09} claim that the SiO and CO outflows are centered on sources $Ab$ and $Aa$, respectively. 
Nevertheless, it is hard to disentangle the powering sources from their molecules maps, since that the separation
between $Aa$ and $Ab$ ($\sim 0.6^{\prime\prime}$) is much smaller than the HPBW of their SiO and CO maps
($\sim 3$\arcsec).

\begin{figure}[t]
\centering
\includegraphics[width=\columnwidth]{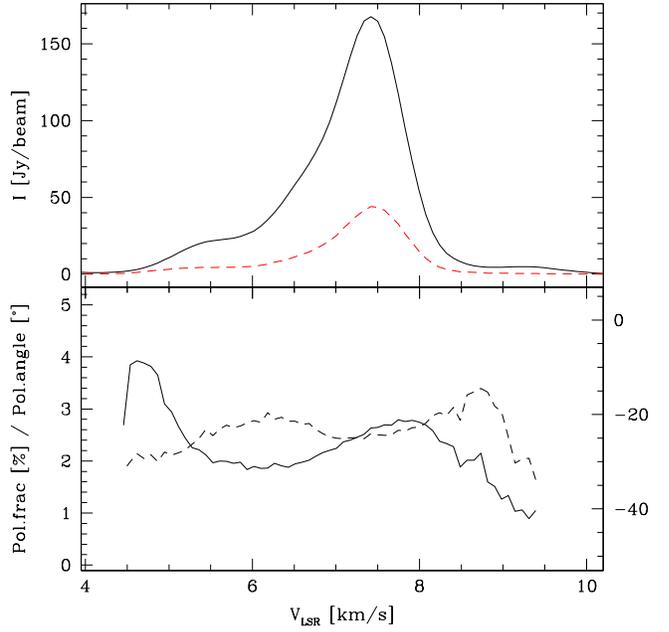}
\caption{Stokes I (black line) 
and linear polarization intensity (multiplied by a factor of 10, red dashed line) spectra of the water 
maser emission (upper panel). The spectra of polarization fraction (black line, left scale) and position angle (dashed line, right scale) are also shown (lower panel).}
\label{stokesIpolfrac}
\end{figure}

Our maser features are also detected at redshifted velocities and lie in a similar direction
as the CO outflow. On the other hand, their positions south of $Aa$ suggest that they may be 
excited by the SiO outflow. In fact, the SE lobes observed for both SiO and CO at redshifted 
velocities \citep[Fig. 6 
and 7 from][]{Rao09} match quite well to the LSR of our highest velocity features (\vlsr $\sim$ 9 \kms). In addition,
VLA 3.7 cm data from \citet{Chandler05} show two sources, A1 and A2, close to $Aa$ (see Fig. \ref{maser_pos}). 
These authors refer to A2 as a protostar which powers a radio jet due to the bipolar shape seen in their
high-resolution maps. On the other hand, source A1 would be an ionized region produced by shocks between this 
flow and nearby dense gas. Therefore, it is possible that the water masers are generated in the interaction regions 
traced by the radio sources and the outflows.

Water masers detections were also reported toward source A through very long baseline (VLBI) 
observations \citep{Imai07}. 
The milliarcsecond resolution data revealed a spot exactly 
over $Aa$  and another one to the SW of it (see Fig. 
\ref{dust_dist}). While the former may have been excited in the circumstellar gas of source 
$Aa$, the latter is likely created by shocks in the E--W outflow. The redshifted lobe of this emission has an
open shell structure, therefore masers off from the outflow main axis are 
expected.

\section{The line-of-sight magnetic field strength in I16293}
\label{field_stre_sec}

From the Zeeman splitting formalism, the magnetic field strength can be correlated to the fraction of circular 
polarization by \citep[see, for instance,][]{Fiebig89}
\begin{eqnarray}
\label{circpol_eq}
P_V & = & (V_{max} - V_{min})/I_{max} \\
\label{bstren_eq}
& = & 2 \times A_{F-F^{\prime}} \times (B \cos \theta)/\Delta V_{I}, 
\end{eqnarray}
where the $A_{F-F^{\prime}}$ coefficient depends on the maser rotational levels $F$ and $F^{\prime}$, the 
intrinsic thermal linewidth $\Delta\nu_{th}$ and the maser saturation degree, while $\Delta v_{I}$ is the $FWHM$ 
of the total power spectrum. 
For the $A_{F-F^{\prime}}$ coefficient, 
we adopted the range of 0.012 to 0.018, since these values are consistent with models and observations
\citep{Nedoluha92,Vlemmings02}.
The water maser spectrum is a blend of  
different velocity components, so we adopt a linewidth ranging from 0.75~\kms, a typical value found in other regions  
\citep{Vlemmings06b}, to 1.0 \kms, the maximum estimated value in I16293 from our spectral profile. For these values, we find that $B_{LOS}$ ranges
from $\sim -94$ to $\sim -188$ mG.  The negative signal is inferred from the Stokes V shape, meaning 
that the field is pointing toward the observer. For narrower linewidths ($\sim 0.75$ \kms), which is more
realistic for resolved water maser lines, the field strength ranges from $\sim -94$ to $\sim -141$ mG (see Table
\ref{b_calc}). We note, however, that given that the linear polarization fraction is about 3\%, this 
maser is likely unsaturated and the field strength determination is a fair approximation. 

\begin{figure}[t]
\centering
\includegraphics[width=\columnwidth]{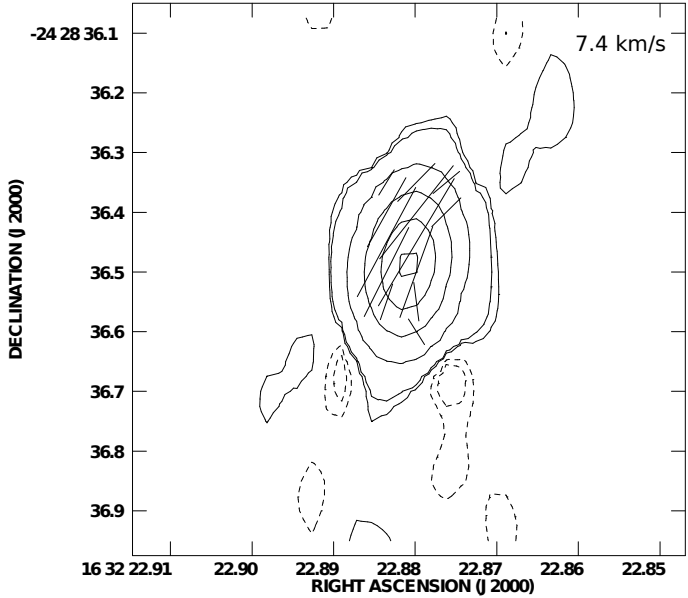}
\caption{Distribution of \hho\ linear polarization vectors in the 
brightest emission channel (\vlsr $\simeq 7.4$ \kms). Contours are -50, -30, 30, 50, 500, 3 $\times$ 
10$^{3}$, 1 $\times$ 10$^{4}$, 2 $\times$ 10$^{4}$ $\times$ 8 \mJybm. 
Only polarization vectors whose $P > 1\%$ are plotted. Vectors are sampled as 1/2 of a 
beam.}
\label{maser_pol}
\end{figure}

Fig. \ref{maser_pol} shows the linear polarization vectors at the velocity channel with the brightest emission. 
There is a degeneracy between the position angle of the linear polarization vectors and the magnetic field direction 
projected in the plane-of-sky (POS). The position angle of the polarization will be parallel or perpendicular to the 
magnetic field in the POS for $\theta > \theta_{{\mathrm crit}}$ or  $\theta < \theta_{{\mathrm crit}} = 55\degr$, 
respectively \citep{Goldreich73}, where $\theta$ is the angle between $B$ and the maser propagation direction
and $\theta_{{\mathrm crit}}$ is the so called Van Vleck angle. Given that the
linear polarization fraction $P_{l}$ depends on $\theta$ and the maser saturation level, we could in principle 
constrain the value of $\theta$ from our data and solve for this ambiguity. We used a non-LTE radiative transfer model in order to fit the observed intensity $I$ and polarization fraction $P_{l}$. However, since our line is a blend of 
features spatially unresolved,  
we were unable to determine $\theta$. Although observations at more extended configurations are needed to achieve this goal,
we can at least claim that the POS field topology is quite ordered in both cases, i. e., it would 
remain ordered if rotated by 90$\degr$.

\begin{figure}[t]
\centering
\includegraphics[width=\columnwidth]{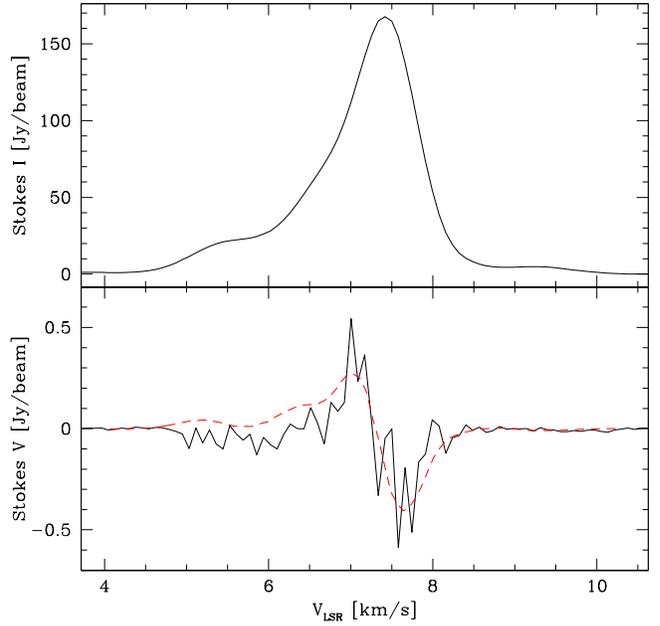}
\caption{Stokes I (upper panel) and Stokes V (lower panel) spectra of the water maser emission. 
The red dashed line is the scaled derivative of total power $I$ in order to guide the eye.}
\label{stokesV}
\end{figure}

According to the submm data of 
\cite{Rao09}, the direction of the dust polarization vectors associated with source A averages to 95$\degr$, 
which implies that the POS magnetic field lines have a PA of 5$\degr$ with respect to the North direction. 
Assuming that the 22 GHz polarization vectors in Fig. \ref{maser_pol} are parallel to the POS magnetic field, the 
main direction of the field lines is $-23\degr$ (or its supplement, 157$\degr$) with respect to the North direction, 
as reported in Section \ref{sec_polem}. On the other hand, if our polarization vectors are perpendicular to the 
POS field, their main direction would then be 67$\degr$. Although two distinct scales are being 
compared, it is interesting to point out that in both cases there is clearly a change in the direction of the
field lines in the small scales traced by our VLA data. In the case of parallel field lines and polarization vectors, 
such a change could be produced by the interaction of the SiO outflow and the magnetic field, 
given that the field lines lie almost in the same direction of the outflow (PA $\sim 145\degr$, Fig. \ref{dust_dist}).

\section{Preshock magnetic fields and post-shock densities}
\label{core_evolution}

The only previous estimation of the magnetic field strength toward I16293 was performed 
by \citet{Claussen03}, who used VLBI observations of water masers 
to estimate $B_{LOS} \sim - 40$ mG.
Their maser spots appear associated with the A1 cm source (Fig. \ref{maser_pos}), with a projected
separation of $\sim 20$ AU at \vlsr $\sim 2.5$ \kms (Claussen, private communication). Since this LSR velocity is similar to 
the core ambient velocity, it is unlike that this emission traces the same region observed with our VLA data.
In any case, our work is then one of the first determinations of the field strength in a low-mass young stellar 
object at densities larger than 10$^{8}$ cm$^{-3}$. As discussed in \S~\ref{mas_outf_sec}, the observed
maser emission is excited in zones of compressed gas produced by shocks between the outflows and the 
ambient gas. In an ionized medium, shocks compress the gas by a factor $m_{A}$, the Alfv\'enic 
Mach number defined as $v_{s}/v_{A}$, where $v_{s}$ is the shock velocity and $v_{A}$ the Alfv\'en speed. 
The magnetic field $B_{0}$ in the pre-shock zone is also enhanced by a factor $m_{A}$ in the shocked gas so that
$B_{s} \sim m_{A} \times B_{0}$. Since $v_{A}$ varies with $B_{0} (n)^{-1/2}$, then $B_{s}$ 
is proportional to $v_{s} (n)^{1/2}$, where $n$ is the volume density of the pre-shock region.
Rearranging this expression, the pre-shock density can be written as \citep{Kaufman96}: 

\begin{figure}[t]
\centering
\includegraphics[width=\columnwidth]{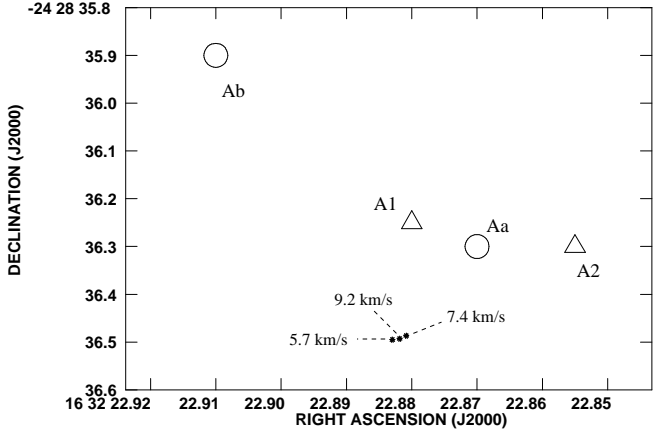}
\caption{Possible independent maser features (stars) as derived by gaussian fits. The accuracy in
the position of each maser feature is $\sim 2$ mas, while the positions of the submm and cm sources 
are accurate within $\sim 25$ mas and $\sim 10$ mas, respectively. The numeric labels
are the maser velocities with respect to the Local Standard of Rest (LSR).
The 3.7 cm sources A1 and A2 (triangles) and the submm
sources $Aa$ and $Ab$ (circles) are also shown \citep{Chandler05}. The sizes of each symbol are not proportional
to their deconvolved sizes or their intensities.}
\label{maser_pos}
\end{figure}

\begin{equation}
\label{preshockn_eq}
n_{0} = 1.6 \times 10^6 \left(\frac{B_{s}}{\mathrm{mG}}\right)^{2} \left(\frac{v_{s}}{\mathrm{km~s^{-1}}}\right)^{-2}
\mathrm{cm}^{-3}.
\end{equation}
The LOS magnetic field strength calculated in \S~\ref{field_stre_sec} can be rescaled in order to obtain
the total field strength $B_{tot} = B_{\mathrm{s}}$ in the shocked zone. According to 
\citet{Crutcher99}, the squared LOS magnetic field $ \langle B_{\mathrm{LOS}} \rangle^{2}$ averaged over several 
lines of sight in the Galaxy is $\sim B_{\mathrm{tot}}^{2}/3$. The shock velocities are retrieved from submm
water emission detected in shocked zones around I16293 \citep{Risto05}. These data were modeled as C-shocks
\citep[low velocity, non-dissociative shocks,][]{Kaufman96} with shock 
velocities between $12-15$ \kms. Applying those
values to equation \ref{preshockn_eq}, we obtain pre-shock densities of $\sim 4.2^{+2.3}_{-2.4} \times 10^8$ \cmt,
which is consistent with the typical densities where water masers are eventually pumped. The upper/lower limits in 
the number density refer to the $B_{s}$ determinations for the two limit $A_{F-F^{\prime}}$ cases 
(Table \ref{b_calc}).

\begin{table}[t]
\caption{$B$-field strength calculation from Eq. \ref{bstren_eq} (assuming $\Delta V_{I} = 0.75$ \kms)}             
\label{b_calc}      
\centering                          
\begin{tabular}{c c c c}        
\hline\hline                 
 & $A_{F-F^{\prime}} = 0.012$ & $A_{F-F^{\prime}} = 0.015$ & $A_{F-F^{\prime}} = 0.018$ \\    
\hline                        
 $B_{\mathrm{LOS}}$ (mG) & $-141$ & $-113$ & $-94$ \\      
\hline                                   
\end{tabular}
\end{table}

If the magnetic field plays a dominant role over the outflow dynamics, the magnetic pressure 
should be similar to the shock ram pressure, i.~e., 
\begin{equation}
\label{magram}
B_{s}^2/8\pi = \rho_{0}v_{s}, 
\end{equation}
where $\rho_{0}$ is the pre-shock 
density. For the field strength $B_s$ determined in the previous section and shock velocities of \cite{Risto05}, we find a mean pre-shock 
density of $\sim 5.1 \times 10^8$ \cmt, which is consistent with the pre-shock densities calculated from equation 
\ref{preshockn_eq} and in line with a magnetically controlled outflow evolution. 

Assuming magnetic flux-freezing, the compression of an ordered magnetic field ($B_0$) amplifies the field 
strength in proportion to the gas density as follows

\begin{equation}
\label{b_compres}
\frac{B_0}{n_0} = \frac{B_s}{n_s}
\end{equation}
where $n_s$ is the post-shock density. If we consider that the shock compresses the well established field geometry in the submm maps of \citet{Rao09}, which predict a (scaled) $B_{0} \simeq 9$ mG at densities $n_0 \simeq 4.9 
\times 10^7$ \cmt, and using the post-shock field intensity $B_s$ calculated from our data, 
we find post-shock densities of $\sim 1.3 
\times 10^9$ \cmt, which is expected for effective water maser pumping. In a more conservative approach,
we rescale the magnetic field strength of \citet{Rao09} using $B \propto n^\kappa$, where $\kappa$ may assume 
distinct vales depending on the dynamics of the core: $\kappa = 0.5$ for a cloud evolution controlled by the 
magnetic field \citep{Crutcher99}, $\kappa = 0.47$, magnetic evolution as modeled by \citet{Fiedler93} and
$\kappa = 0.67$, for a free-fall cloud collapse. For the three cases, the pre-shock magnetic field in the
calculated pre-shock densities ranges between $\sim$ 24 and 36 mG. Applying again equation \ref{b_compres}, we
obtain post-shock densities ranging 
between $\sim 2-5 \times 10^9$ \cmt, again consistent with typical
\hho maser densities, but unable to discern in which dynamical regime, if magnetic or turbulent, the core evolves.

The \hho\ maser emission of I16293 is strong ($> 200$ Jy) and stable over a few weeks but highly 
variable over months, as shown by
surveys performed toward this source \citep{Wilking87,Claussen96,Furuya03}. We are then encouraged to carry 
out VLBI observations in the future. 
With these observations, the blended maser components found in 
this work will be likely resolved both spatially and spectroscopically. We will then be able to apply radiative
transfer models such as in
\citet{Vlemmings06}  to derive the full 3-D magnetic field properties as 
already done in other star-forming regions \citep[e. g., Cepheus A HW2, ][]{Vlemmings06b}.

\section{Conclusions}
\label{sec_mas_con}

This work reports on the \hho (6$_{16} - 5_{23}$) maser emission observed with the VLA/EVLA toward 
the low-mass source IRAS 16293-2422. This is one of the first estimations of the line-of-sight magnetic
field strength in a low-mass protostar 
at densities larger than 10$^{8}$ cm$^{-3}$. 
From this study, we obtained insights on
the dynamics of this object at such high densities.
Our main conclusions are:

\begin{itemize}

\item We detect strong water maser emission associated with the submm source $Aa$.

\item The maser spectrum has a non-Gaussian profile, indicating the presence of at least three maser features covering a velocity range of $\sim$ 3.5 \kms~distributed
in a linear configuration. Those components are likely excited in zones of compressed gas produced by 
shocks in the outflow activity of I16293.

\item The obtained Stokes V spectrum is consistent with Zeeman emission. The mean LOS magnetic field strength 
is $ \sim 113$~mG.
The POS field topology retrieved from the linear polarimetry shows an ordered pattern consistent with larger scales 
field morphologies.

\item The post- and pre-shock densities calculated from our field strength estimation are consistent with the
typical densities expected for \hho maser pumping.

\item The dynamics of the outflow evolution in these sources are likely regulated by the magnetic field, since that
the magnetic pressure in the shocked gas is similar to the pre-shock ram pressure.   

\end{itemize}

\begin{acknowledgements}
The authors would like to thank the anonymous referee for a constructive report that helped to improve the paper.
We are also grateful to Gabriele Surcis for running the radiative transfer models on the maser data.
This research was partially supported by the Deutsche Forschungsgemeinschaft (DFG) through the Emmy Noether 
Research grant VL 61/3-1 and through SFB 956.
FOA, JMG and JMT also acknowledge support from MICINN (Spain) AYA2011-30228-C03 and
AGAUR (Catalonia) 2009SGR1172 grants. 
\end{acknowledgements}

\bibliographystyle{aa}
\bibliography{refalves}

\end{document}